\documentclass[pra,aps,twocolumn,showpacs]{revtex4}
\usepackage{graphicx}
\usepackage{dcolumn}

\begin{document}
\def\be{\begin{equation}}
\def\ee{\end{equation}}
\def\bearr{\begin{eqnarray}}
\def\eearr{\end{eqnarray}}
\def\la{\langle}
\def\ra{\rangle}
\def\l{\left}
\def\r{\right}

\title{Multibranch Bogoliubov-Bloch spectrum of a 
cigar shaped Bose condensate in an optical lattice}

\author{Tarun Kanti Ghosh and K. Machida}
\affiliation
{Department of Physics, Okayama University, Okayama 700-8530, Japan}

\date{\today}

\begin{abstract}
We study properties of excited states of an array of
weakly coupled quasi-two-dimensional Bose condensates 
by using the hydrodynamic theory.
The spectrum of the axial excited states strongly depends 
on the coupling among the various discrete radial modes 
in a given symmetry.
By including mode-coupling within a given symmetry, the 
complete excitation spectrum of axial quasiparticles with
various discrete radial nodes
are presented. 
A single parameter which
determines the strength of the mode coupling is identified.
The excitation spectrum in the zero angular momentum sector can be 
observed by using the Bragg scattering experiments.
\end{abstract}

\maketitle

\section{Introduction}
The experimental realization of optical lattices \cite{op} is 
stimulating new perspectives in the study of cold bosons.
Optical lattices have enabled us to observe quantum phenomena
such as number squeezing \cite{op1}, collapses and revivals \cite{op2} and 
the diffraction of matter waves \cite{op3}. Apart from these examples, BEC
in optical lattices are particularly promising physical systems to 
study the superfluid properties of Bose gases \cite{op4,op5}.
The Bose-Hubbard model has been realized and the 
quantum phase transition from superfluid to a Mott insulator state 
was indeed observed experimentally \cite{bh1,bh2,bh3}.
It was predicted that for deep optical lattices the condensate superflow 
can be lost not only by energetic instability but also by dynamical instability 
\cite{inst1,inst2,inst3}. The dynamic instability was
verified by the experiments \cite{fallani}.
In a seminal work by Kramer {\em et al.} \cite{kramer}, they have found the mass 
renormalization
in presence of the optical potential which decreases the value of 
the axial excitation
frequencies. These discrete axial excitation frequencies are 
experimentally verified \cite{kramer1}.
There are several theoretical calculations
\cite{molmer,java,pethick,taylor} for the sound
velocity in a quasi-1D Bose gas placed in an 1D optical lattice.

The axial excitations of a cigar shaped condensate can be
divided into two regimes: i) short wavelength excitations
where wavelength is much smaller than the axial size, ii)
long wavelength excitations where wavelength is equal or
larger than the axial size of the system. 
In the former case, these excitations can be classified with
a continuous wave vector $k$. However, the finite transverse
size of the condensate also produces a discreteness of the
spectrum. The short wavelength
axial phonons with different number of discrete radial modes 
of a cigar shaped condensates (without optical lattice) 
give rise to the multi-branch Bogoliubov spectrum (MBS)
\cite{mbs1,fedichev}.
The MBS was observed in a Bragg spectroscopy with a long duration of 
the Bragg pulses \cite{mbs2}. 
An array of weakly coupled quasi-two-dimensional condensates 
can be created by applying a relatively strong one-dimensional 
optical lattices to an ordinary three dimensional cigar shaped 
condensate along the symmetry axis. In presence of the periodic
lattices, the MBS can be called as multibranch Bogoliubov-Bloch
spectrum (MBBS).
It is useful to study the MBBS 
in view of the possibility of the experimental verification.
It should be noted that all the modes in a given angular 
momentum sector are coupled among themselves for
any finite value of the axial momentum. For example,
when we excite the system to study the sound propagation along
the symmetry axis, this 
perturbation inherently excites all other low energy transverse 
modes having zero angular momentum. Therefore,
all modes are coupled with each other in the same angular momentum sector. 
Martikainen and Stoof \cite{mbs3} have studied the 
MBBS only for monopole and lowest energy quadrupole modes without considering
the coupling among the various modes within a given symmetry
by means of time-dependent Gaussian variational ansatz.
Later, Martikainen and Stoof \cite{mbs4} have calculated 
the spectrum of the  phonon and the monopole modes
by considering only the coupling between the phonon 
and the breathing modes. But it is noted that
the sound mode is coupled not only with the breathing mode
but also with other low energy modes having zero angular momentum.
Similarly, the lowest energy quadrupole mode is also coupled
with other low energy quadrupole modes.
There is a lack of complete study on the MBBS in this
system.
For complete and correct description of MBBS
we have to consider the couplings among all low energy modes in the same
angular momentum sector. In our discretize hydrodynamic description,
the couplings among all the modes in the same angular momentum sector 
are included naturally and we will see in the next section. 

In this work we study the excitations in a stack of weakly coupled
quasi-two-dimensional condensates. 
The multibranch Bogoliubov-Bloch spectrum of such system is presented
by using the hydrodynamic theory.
The MBBS strongly depends on the coupling between the inhomogeneous
density in the radial plane and the density modulation along the symmetry axis.
Note that one can study only the spectrum of sound, monopole and quadrupole modes
without considering the mode coupling completely 
by using the time-dependent Gaussian variational method. 
Our discretize hydrodynamic method presented in this paper goes beyond 
the time-dependent variational method. 
In principle, we can calculate all low energy spectrum by including the
mode coupling in a given angular momentum sector as long as the 
excitation energies are less than the chemical potential. 
We find that the multibranch Bogoliubov-Bloch spectrum
changes due to presence of the mode-coupling within a given angular
momentum symmetry. Therefore, the mode-coupling should be taken into
account while calculating the spectrum correctly.

This paper is organized as follows. In Sec.II, we consider an array
of weakly coupled quasi-two-dimensional Bose condensates. Using the discretize
hydrodynamic theory, we calculate the multibranch Bogoliubov-Bloch spectrum
by including the mode coupling within a given symmetry.
We give a brief summary and conclusions in Sec. III.

\section{MBBS of a non-rotating array of Bose condensates}
We assume that the bosonic atoms, at $ T=0$, are trapped by an
external potential given by the sum of a harmonic trap
and a stationary optical potential modulated along the 
$z$ axis. The Gross-Pitaevskii energy functional can be written as
\bearr
E_0 & = & \int dV \psi^{\dag}(r,z)[- \frac{\hbar^2}{2M} {\bf \nabla}^2 + 
V_{ho}(r,z) 
\nonumber \\ & + &  
\frac{g}{2} |\psi(r,z)|^2 + V_{op}(z)] \psi(r,z).
\eearr
Here, $ V_{ho}(r,z) = \frac{M}{2} (\omega_r^2 r^2 + \omega_z^2 z^2) $ is the
harmonic trap potential and
$ V_{op}(z) = s E_r \sin^2(qz) $ is the optical potential where $ E_r = 
\frac{\hbar^2 q^2}{2M} $
is the recoil energy, $ s$ is the dimensionless parameter determining the
laser intensity and $ q $ is the wave vector of the laser beam. Also,
$ g = \frac{4 \pi a \hbar^2}{M} $ is the strength of the two-body interaction energy, 
where $a$ is the two-body scattering length.
We also assumed that $ \omega_r >> \omega_z $ so that it makes
a long cigar shaped trap. The minima of the optical potential are located
at the points $ z_j = j \pi/q = j d $, where $d = \pi/q $ is the
lattice size along the $z$-axis. Around these minima, 
$ V_{op}(z) \sim M/2 \omega_s^2 (z-z_j)^2 $, 
where the layer trap frequency is $ \omega_s = \sqrt{s} \hbar q^2/M $. 
In the usual experiments, the well trap frequency is larger than than 
the axial harmonic frequency, $  \omega_s >> \omega_z $.
Therefore, we can also ignore the harmonic
potential along the $z$-axis since the deep optical lattice
dominates over the harmonic potential along the $z$-axis.

The strong laser intensity will give rise to an array
of several quasi-two-dimensional condensates. Because of the quantum 
tunneling, the overlap between the wave functions between two
consecutive layers can be sufficient to ensure full coherence.
If the tunneling is too small, the strong phase fluctuations 
will destroy the coherence and lead to a new quantum state, 
namely Mott insulator state. 

In the presence of coherence among the layers it is natural 
to take the ansatz for the wave function as 
\be 
\psi(x,y,z) = \sum_j \psi_j(x,y) f(z-z_j).
\ee
Here, $  \psi_j(x,y) $ is the wave function of the two-dimensional
condensate at the site $j$ and $ f(z-z_j) $ is a 
localized function at $j$-th site.
The localized function can be
written as
\be \label{ansatz}
f(z-z_j) = (\frac{M \omega_s}{\pi \hbar})^{1/4} 
e^{-\frac{M \omega_s}{2 \hbar} (z-z_j)^2}.
\ee
Substituting the above ansatz into the energy functional and
considering only the nearest-neighbor interactions, one can 
get the following energy functional:
\bearr \label{main}
E_0 & = & \sum_j \int dx dy [-\frac{\hbar^2}{2M} 
\psi_j^{\dag} {\bf \nabla}_r^2 \psi_j
+ V_{ho} |\psi_j|^2]
\nonumber \\ & + &
\frac{ g_{\rm 2D}}{2} \sum_j \int dx dy \psi_j^{\dag} \psi_j^{\dag} \psi_j \psi_j
\nonumber \\ & - &
J \sum_{j,\delta = \pm 1} \int dx dy [\psi_{j+\delta}^{\dag} \psi_j +  \psi_j^{\dag} 
\psi_{j+\delta}].
\eearr
Here, $ J $ is the strength of the Josephson coupling 
between adjacent layers which is 
given as
\bearr
J & = & -  \int dz   f(z) [- \frac{\hbar^2}{2M}   \nabla_z^2 +
V_{op}(z) ] f(z + d) \nonumber \\
& \sim & \hbar \omega_r (\frac{ \pi a_r}{\sqrt{2} \lambda})^2 (\pi^2 - 4)s 
e^{-\frac{\pi^2 \sqrt{s}}{4}},
\eearr
where $ a_r = \sqrt{\frac{\hbar}{M \omega_r}} $. 
Also, the strength of the effective on-site interaction energy is
$  g_{\rm 2D} = g \int dz |f_0(z)|^4 = 4 \sqrt{\frac{\pi}{2}}
\frac{\hbar^2}{M}(\frac{a}{a_s})$,
where $ a_s = \sqrt{\hbar/M \omega_s} $.
Eq. (\ref{main}) shows that each layer $j$ is coupled with the
nearest-neighbor layers $ j\pm 1$ through the tunneling energy $J$.
The axial dimension appears through the Josephson coupling between 
two adjacent layers. 
The Hamiltonian corresponding to the above energy functional is
similar to an effective 1D Bose-Hubbard Hamiltonian in which each lattice
site is replaced by a layer with radial confinement.

The Heisenberg equation of motion for the bosonic order parameter 
is
\be
i \hbar \dot \psi_j   =  [-\frac{\hbar^2}{2M} {\bf \nabla}_r^2
+ V_{ho} +  g_{\rm 2D} \psi_j^{\dag} \psi_j] \psi_j 
-J (\psi_{j-1} + \psi_{j+1}).
\ee
Using the phase-density representation of the bosonic field operator as
$ \psi_j = \sqrt{n_j} e^{i\theta_j} $ and neglecting the quantum
pressure term, one can get the following equations
of motion for the density and phase:
\bearr
\dot n_j  & = & - \frac{\hbar}{M} {\bf \nabla}_r \cdot (n_j {\bf \nabla}_r \theta_j)
+  \frac{2J}{ \hbar} [ \sqrt{n_j n_{j-1}} \sin(\theta_j - \theta_{j-1})
\nonumber \\ & - & \sqrt{n_j n_{j+1}} \sin(\theta_{j+1} - \theta_{j})],
\eearr
and
\bearr
\hbar \dot \theta_j & = &
- \frac{\hbar^2}{2M} ({\bf \nabla}_r \theta_j)^2 
+ J [ \sqrt{\frac{n_{j+1}}{n_j}} \cos(\theta_{j+1} - \theta_{j})
\nonumber \\
& + &  \sqrt{\frac{n_{j-1}}{n_j}} \cos(\theta_{j} - \theta_{j-1})]
- V_{ho} -  g_{\rm 2D} n_j.
\eearr
Here, $ ``\cdot" $ represents the time derivative. In equilibrium, the condensate
density at each layer is $ n_{0}(r) = \frac{\mu_0 - V_{ho}(r)}{ g_{\rm 2D} } $,
where we have neglected the effect of the tunneling energy $ J$ since it is very
small in the deep optical lattice regime. 
Also, $ \mu_0 = \hbar \omega_r \sqrt{\sqrt{8/\pi} \frac{Na}{a_s}} $ is the 
chemical potential at each layer, where $N$ is the number of atoms
at each layer.
In this system, we have two energy scales: the chemical potential of
each layer $ \mu_0 \sim s^{1/8}$ which is associated
with the radial plane and the tunneling energy
$ J \sim s e^{-\sqrt{s}}$ which is associated with the density modulation
along the $z$-axis.
The strength of the chemical potential can be enhanced by 
increasing the lattice depth or by increasing
the number of atoms. The tunneling energy $J$ decreases with the increasing
of the lattice depth.

We linearize the hydrodynamic equations around the equilibrium state, 
as $ n_j = n_0 + \delta n_j $ and
$ \theta_j = \delta \theta_j $. The equations of motion for the density and phase
fluctuation becomes
\bearr
\delta \dot n_j & =  &- \frac{\hbar}{M} {\bf \nabla}_r \cdot [n_{0}(r) {\bf \nabla}_r \delta 
\theta_j] 
\nonumber \\
& + &  \frac{2J}{ \hbar} n_{0}(r) [2 \delta \theta_j -  \delta \theta_{j-1} - \delta 
\theta_{j+1}]
\eearr
and
\be \label{vfluc}
\hbar \delta \dot \theta_j = -  g_{\rm 2D} \delta n_j - \frac{J}{2n_0(r)}
[2\delta n_j - \delta n_{j-1} - \delta n_{j+1}].
\ee
Note that the second term of the right hand side of Eq. (\ref{vfluc})
is proportional to the small parameter $J$ and inversely proportional
to the large parameter $ n_0(r =0) = \mu_0/ g_{\rm 2D} $. 
Therefore, we can neglect the term which is proportional to the
$\frac{J}{2n_0(r)} $. After some algebra, we get 
second order equation of motion for
the density fluctuation as 

\bearr \label{density}
\delta \ddot n_j & = & \frac{ g_{\rm 2D}}{M} {\bf \nabla}_r \cdot [n_{0}(r) {\bf \nabla}_r 
\delta n_j]
\nonumber \\
& - &
\frac{2 J g_{\rm 2D}}{\hbar^2} n_{0}(r) [2  \delta n_j - \delta n_{j-1} - \delta 
n_{j+1}].
\eearr 
The above equation tells us that the density fluctuation at each layer $j$ is 
coupled with the nearest-neighbor layers $ j \pm 1 $.
We seek the normal mode solutions of the density fluctuations at layer $j$ 
in the following form:

\be 
\delta n_j  = \delta n(r) e^{i[jkd-\omega_l(k)t]}.
\ee
Here, $ k $ is Bloch wave vector (quasi-momentum) of the excitations.
The Bloch wave vector $p$ which is associated with the velocity
of the condensate in the optical lattice is set to zero.
 
Substituting the above equation into Eq. (\ref{density}), we get 
\be \label{density1}
-\omega_l^2(k) \delta n  = \frac{ g_{\rm 2D}}{M} {\bf \nabla}_r \cdot (n_{0} 
{\bf \nabla}_r \delta n)
-\frac{8 J g_{\rm 2D}}{\hbar^2} n_{0}(r) \sin^2(kd/2) \delta n,
\ee
where $ l $ is a set of two quantum numbers: radial quantum number, $ n_r $
and the angular quantum number, $m$. The parameter $J \mu $ in front of the
$ \sin^2(kd) $ term determines the strength of the coupling between the
inhomogeneous density in the radial plane and the density modulation along
the $z$ axis.
  
For $ k =0 $, the solutions are known exactly and analytically
\cite{graham}.
The energy spectrum and the normalized eigen functions, respectively, are given as,
$ \omega_l^2 = \omega_r^2[|m| + 2 n_r(n_r +|m| +1)] $ and
\be
\delta n(r,\phi) = \frac{(1+2n_r+|m|)^{1/2}}{(\pi R_0^2)^{1/2}} \tilde r^{|m|} 
P_{n_r}^{(|m|,0)}(1-2 \tilde r^2) e^{i m \phi}.
\ee
Here, $ P_n^{(a,b)}(x) $ is the Jacobi polynomial of order $n$ and $ \phi $ is the 
polar angle. The radius of each condensate layer is $ R_0 = 2 \mu_0/M \omega_r^2 $ and
$ \tilde r = r/R_0 $ is the dimensionless variable. 

The solution of Eq. (\ref{density1}) can be obtained for arbitrary value of $k$ by
numerical diagonalization. 
For $ k \neq 0 $, we can expand the density fluctuations as
\be
\delta n(r) = \sum_l b_l \delta n_{l}(r,\phi).
\ee 
Substituting the above expansion into Eq.(\ref{density1}), we obtain,
\bearr \label{density2}
0 & = & [\tilde \omega_l^2 - [|m| + 2 n_r(n_r +|m| +1)] 
\nonumber \\ & - &
B_0 \sin^2(kd/2)] b_l + 
B_0 \sin^2(kd/2) \sum_{l^{\prime}} M_{ll^{\prime}} b_{l^{\prime}},
\eearr 
where $ \tilde \omega_l = \omega_l/\omega_r $ and the dimensionless parameter
$B_0$ is defined as
\be \label{imp}
B_0 = \frac{8 J \mu_0}{\hbar^2 \omega_r^2}. 
\ee 
The matrix element $ M_{ll^{\prime}} $ is given by
\bearr \label{matrix}
M_{ll^{\prime}} & = & \frac{(1+2n_r+|m|)}{\pi} \int d^2 \tilde r 
\tilde r^{2 + |m| + |m^{\prime}|} e^{i(m - m^{\prime}) \phi} 
\nonumber \\ & \times & P_{n_r^{\prime}}^{(|m^{\prime}|,0)}(1-2 \tilde r^2)
P_{n_r}^{(|m|,0)}(1-2 \tilde r^2).
\eearr
The above eigenvalue problem is block diagonal with no overlap
between the subspaces of different angular momentum, so
that the solutions to Eq.(\ref{density2}) can be obtained separately in
each angular momentum subspace. We can obtain all low energy
multibranch Bogoliubov-Bloch spectrum from Eq. (\ref{density2})
which is our main result. 
Equations.(\ref{density2}) and (\ref{matrix}) show that the spectrum depends 
on average over the radial coordinate and the coupling among 
the modes within a given angular momentum symmetry for any finite value of $k$ .
Particularly, the couplings among all other modes are important for large
values of $kd $ and $ B_0$. 
It is interesting to note that the curvature of a mode spectrum depends
on a single parameter $ B_0 $ which is defined in Eq. (\ref{imp}).
The parameter $B_0$ can remain unchanged by changing values of the
$J$ and $\mu_0 $ in a various combination.   
Therefore, the curvatures of the spectrum of a given mode for various combinations
of $J$ and $\mu_0 $ with fixed $B_0$ are the same.   

Before presenting the exact numerical results, we make some approximation
for a quantitative discussions.
If we neglect the couplings among  all other modes in the $m=0$ sector by setting
$ l^{\prime} = (n_r, 0) $ in Eqs. (\ref{density2}) and (\ref{matrix}), one 
can easily get following spectrum: 
\be \label{per}
\tilde \omega_{n_r}^2 = 2n_r(n_r+1) + (1-M_{n_r,n_r}) B_0 \sin^2(kd/2).
\ee
The above equation can also be obtained by using first-order perturbation
theory to Eq. (\ref{density1}).
In the limit of long wavelength, the $ n_r = 0 $ mode is phonon-like with a sound 
velocity
$c_0 = \sqrt{\frac{\mu_0}{2M^*}} $, where $ M^* = \frac{\hbar^2 }{ 2 J d^2} $
is the effective mass of the atoms in the optical potential. 
This sound velocity exactly matches with the result obtained in
Ref. \cite{kramer} and is similar to the result 
obtained without optical potential \cite{mbs1}. 
This sound velocity is smaller by a factor of $ \sqrt{2} $ with
respect to the sound velocity
obtained previously \cite{molmer,java,pethick,taylor} for quasi-1D Bose
gas placed in an optical potential. This is due to the effect of the average
over the radial variable which can be seen from Eqs. (\ref{density2}) and 
(\ref{matrix}). 

In Fig.1 we show few low-energy multibranch Bogoliubov-Bloch spectrum 
in the $ m = 0 $ sector as
a function of $kd$ by solving the matrix Eq. (\ref{density2}).

\begin{figure}[ht]
\includegraphics[width=8.0cm]{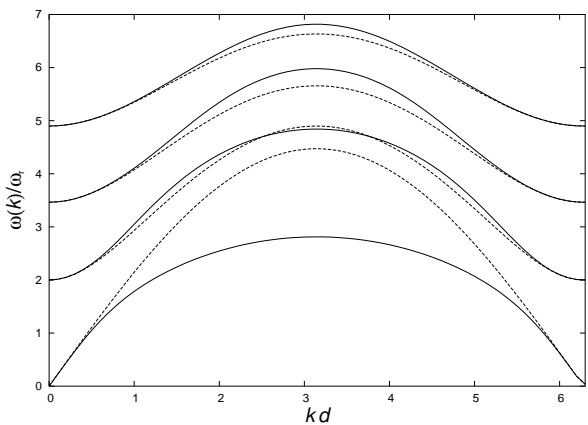}
\caption{Plots of the low-energy Bogoliubov-Bloch modes in 
the $m=0$ sector.
Here, $ J =0.1 \hbar \omega_r $ and $ \mu_0 = 50 \hbar \omega_r $.
Solid and Dashed lines are obtained from 
Eq. (\ref{density2}) and Eq. (\ref{per}), respectively}.
\end{figure}

The lowest branch corresponds to the Bogoliubov-Bloch axial 
mode with no radial nodes. This mode has the usual form like 
$ \omega_r = c_s k $ at low momenta, where $c_s $ is the real 
sound velocity. Note that $ c_s \lesssim c_0 $ which implies 
that the dispersion relations are modified due to the coupling 
among all other modes. The changes in the spectrum is clearly
visible in the central part of the Brillouin zone. 
This is due to the fact that the mode coupling is strong 
enough in the central part of the Brillouin zone
due to the particular nature of the $k$-dependent part (see Eq. (\ref{density2})).
The second branch corresponds to one radial node and starts at 
$ 2 \omega_r $ for $k=0$. The breathing mode has the free-particle 
dispersion relation and it can be written in terms of the
effective mass ($m_b^* $) of this mode as 
$ \omega_2(k) = 2 \omega_r + \frac{\hbar k^2}{2m_b^*} $. 
Fig.1 shows that the mode-coupling does not affect on the breathing mode spectrum
appreciably. 
The third and fourth lowest energy modes are also given in Fig. 1. 
These modes are also changed in the central part of the Brillouin 
zone due to the mode-coupling.
One could see from Fig.1 that the effective masses of each modes are different.
The group velocity along the $z$ direction deviates from its long-wavelength limit 
when $ kd \sim \pi $. The mode coupling induced by the $ \sin^2(kd) $ 
perturbation in Eq.(\ref{density}) becomes more significant with increasing $k$ and 
has the effect of lowering the sound speed. This coupling is associated with the
interplay of the density modulation along the $z$ direction and the strong
inhomogeneity of the equilibrium density in the radial direction in each plane.
The effective masses are negative when $ kd > \pi $.

The coupling between the transverse quadrupole modes ($m = \pm 2 $) and the
modes in the $m \neq \pm 2 $ sector does not exist since these modes are orthogonal
to each other as it can be seen from Eq. (\ref{matrix}).
However, the lowest energy quadrupole spectrum ($ n_r = 0, m = \pm 2 $) strongly 
depends on other low-energy
quadrupole modes with various discrete radial nodes ($ n_r \neq 0, m = \pm 2 $).
We neglect the couplings among  all other modes in the $ m= \pm 2 $ sector by setting
$ l^{\prime} = (n_r, 2) $ in Eqs. (\ref{density2}) and (\ref{matrix}), then one
can easily get following spectrum:
\be \label{per1}
\tilde \omega_{n_r}^2 = 2 + 2n_r(n_r+3) + (1-M_{n_r,2;n_r,2}) B_0 \sin^2(kd/2).
\ee
 
In Fig.2, we present first two low-energy MBBS for quadrupole modes. 
\begin{figure}[ht]
\includegraphics[width=8.0cm]{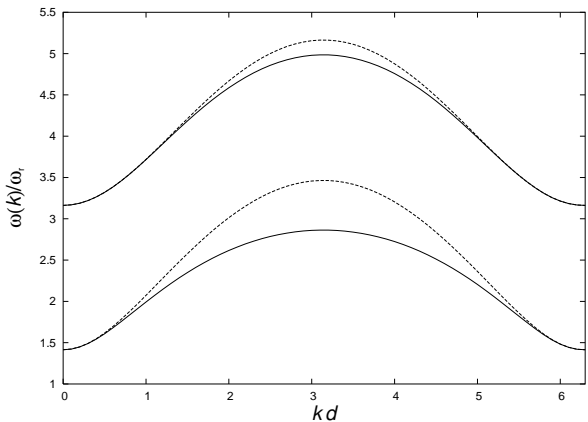}
\caption{Plots of the low-energy Bogoliubov-Bloch modes in
the $m= \pm 2 $ sector.
Here, $ J =0.1 \hbar \omega_r $ and $ \mu_0 = 50 \hbar \omega_r $.
Solid and Dashed lines are obtained from Eqs. (\ref{density2}) and (\ref{per1})}.
\end{figure}
Fig. 2 clearly shows that the mode-coupling reduces the spectrum
also for the quadrupole modes in the central part of the 
Brillouin zone. 

In Ref. \cite{mbs3}, the spectrum for the breathing and 
the lowest energy quadrupole modes are obtained analytically 
within the Gaussian variational analysis. The mode coupling
was not considered in this variational analysis \cite{mbs3}.
In Fig. 3, we compare the spectrum of the breathing and lowest energy 
quadrupole modes obtained from Eq. (\ref{density2}) with those of obtained 
in Ref. \cite{mbs3}. 
It is clear from Fig. 3 that the
mode-coupling reduces the spectrum strongly and it should be taken into account 
for calculating the spectrum correctly. 

\begin{figure}[ht]
\includegraphics[width=8.0cm]{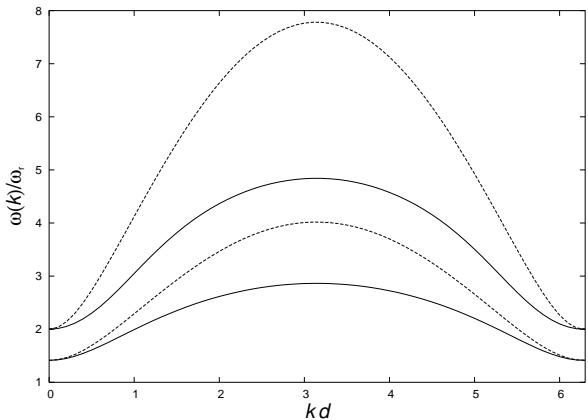}
\caption{Plots of the spectrum of breathing and lowest energy quadrupole modes.
Here, $ J =0.1 \hbar \omega_r $ and $ \mu_0 = 50 \hbar \omega_r $.
Solid and Dashed lines are obtained from
Eq. (\ref{density2}) and Ref. \cite{mbs3}, respectively}.
\end{figure}

\section{Summary and conclusions}

In this work, we have studied excitation energies of the
axial quasiparticles with various discrete radial nodes 
of an array of weakly coupled quasi-two dimensional Bose condensates.
Our discretize hydrodynamic description enables us to produce
correctly all low-energy MBBS by including the mode couplings
among different modes within the same angular momentum sector.
We found that the mode-coupling strongly changes the spectrum.
Therefore, it should be taken into account to calculate such
spectrum correctly. The mode coupling is strong enough in the
central part of the Brillouin zone.
The single parameter $B_0$, defined in Eq. (\ref{imp}),
is identified which is always associated 
with the $k$-dependent part and it scales with the product of two 
energy scales of this system, namely $ J $ and $ \mu_0 $. The parameter
$ B_0 $ is a good measure for determination of the effect of the optical
lattices on the spectrum. 
Particularly, the spectrum for the phonon
and breathing modes can be observed in a Bragg
scattering experiments \cite{phonon} as discussed below. 
The MBBS can be observed in the Bragg scattering experiments 
as the MBS was observed in Ref. \cite{mbs2}. Due
to the axial symmetry, the modes having only zero angular
momentum can be excited in the Bragg scattering experiments.
In the Bragg spectroscopy, the condensate is excited by 
an external moving optical potential 
$ V = V_B(t) \cos(kz-\omega t) $, where $ V_B(t)$ is the
intensity of the Bragg pulses.
This optical potential is created by
using two Bragg pulses with 
approximately parallel polarization, separated by an angle 
$\theta$. The pulses have a frequency difference $\omega $
determined by two acousto-optic modulators. 
The wave-vector $ \bf k$ is adjusted to be along the $z$-axis.
Both the values of $ k$ and $ \omega $ can be tuned by changing 
the angle between two beams and varying their frequency difference. 
For small values of $k$ the system is excited in the
phonon regime and the response is detected by measuring the net
momentum, $ P_{z}(\omega, k) $, imparted to the system by the Bragg
pulses. The multibranch Bogoliubov spectrum is obtained by observing the
locations of the peaks in $ P_{z}(\omega, k) $ for various  values of $k$.
The frequency $ \omega $ must be comparable to radial trap 
frequency $\omega_r$ in order to excite the breathing and
other modes. The duration of the Bragg pulses must 
be larger than the radial trapping period, $ T_B > 2\pi/\omega_r$
in order to have large populations of the radial quasiparticle
states.

\section{acknowledgments}
This work is supported by a fellowship (P04311) of the 
Japan Society for the Promotion of Science.

\end{document}